\documentstyle[epsf,twocolumn,prb,aps]{revtex}
\begin{document}
\draft
\title{Magnetization-plateau state of the $S=3/2$ spin chain 
with single ion anisotropy}
\author{Atsuhiro Kitazawa$^{\dagger}$ and Kiyomi Okamoto$^{\ddagger}$}
\address{$\dagger$Department of Physics, Kyushu University 33, 
Fukuoka 812-8581 Japan \\
$\ddagger$Department of Physics, Tokyo Institute of Technology, 
Oh-okayama, Meguro-ku, Tokyo 152-8551, Japan}
\date{\today}

\maketitle
\begin{abstract}
We reexamine the numerical study of 
the magnetized state of the $S=3/2$ spin chain with single ion 
anisotropy $D(>0)$ for the magnetization $M=M_{S}/3$, 
where $M_{S}$ is the saturation magnetization. 
We find at this magnetization that 
for $D<D_{c1}=0.387$ the system is critical and the magnetization plateau 
does not appear. 
For $D>D_{c1}$, the parameter region is divided into two parts 
$D_{c1}<D<D_{c2}=0.943$ and $D_{c2}<D$. 
In each region, the system is gapful 
and the $M=M_{S}/3$ magnetization plateau 
appears in the magnetization process. 
From our numerical calculation, the intermediate region $D_{c1}<D<D_{c2}$ 
should be characterized by a magnetized valence-bond-solid state. 
\end{abstract}

\section{Introduction}
Recently there has been a considerable interest 
in the magnetization process of one-dimensional quantum spin systems. 
Extending the Lieb-Schultz-Mattis theorem,\cite{LSM} 
Oshikawa {\it et al.}\cite{Oshikawa} gave a necessary condition 
for the appearance of the magnetization plateau 
as $p(S-\langle m\rangle) = \mbox{integer}$, 
where $S$ is the magnitude of the spin, $p$ is the periodicity of the 
magnetic ground state in the thermodynamic limit, 
and $\langle m \rangle$ is the magnetization per site. 

As a simple case, the $S=3/2$ spin chain with single ion anisotropy 
in a magnetic filed 
$h$, 
\begin{equation}
  H = \sum_{j=1}^{L}\mbox{\boldmath$S$}_{j}\cdot\mbox{\boldmath$S$}_{j+1}
    +D \sum_{j=1}^{L}(S_{j}^{z})^{2} -h M
\label{eq:Hamiltonian}
\end{equation}
has been studied. 
Here $\mbox{\boldmath$S$}_{j}$ is the $S=3/2$ spin at the $j$-th site, 
and $M$ is the magnetization $M=\sum_{j}S_{j}^{z}$.
Hereafter, we define the saturation magnetization as $M_{S}(=3L/2)$. 
In this model, a magnetization plateau can appear at $M=M_{S}/3$ ($p=1$). 
For $D=0$ ($S=3/2$ Heisenberg model), 
a magnetization plateau does not appear.\cite{Parkinson} 
On the other hand for large $D$, 
the spin of each site tend to have $S^{z}=1/2$ for $M=M_{S}/3$, 
and a magnetization plateau appears. 
Oshikawa {\it et al.} \cite{Oshikawa} checked that 
a magnetization plateau appears for $D=2$. 
Sakai and Takahashi \cite{Sakai} studied the model 
with the numerical diagonalization calculation using the phenomenological 
renormalization group (PRG) analysis, 
and found that the transition point is $D_{c}=0.93\pm 0.01$. 
They concluded this point as 
the Berezinskii-Kosterlitz-Thouless (BKT) transition \cite{Berezinskii,Kosterlitz73,Kosterlitz74}
point, that is, the plateau and the no-plateau transition point. 
Although the PRG analysis is a powerful method to study the 
second order phase transition, this is not sufficient for 
the BKT transition because of the singular behavior of the energy gap 
and of the logarithmic size dependence of excitation energies. 
Thus we think that it is worthwhile to reexamine this problem 
by use of another numerical approach. 

The organization of this paper is as follows. 
In the next section, we consider an effective continuum model in order to 
investigate the mechanism of the appearance of the $M=M_{S}/3$ magnetization. 
In section 3, we show the numerical method and the results 
with the numerical diagonalization data. 
To study the system, we firstly consider the finite-size behavior 
with conformal field theory and the renormalization group analysis. 
Next, using them, we determine the boundary of the plateau and the no-plateau 
regions, and also check the critical behavior. 
Last section is devoted to summary and discussion. 

\section{Effective continuum model}
For the magnetization $M=M_{S}/3$, 
the effective continuum model describing 
the low energy physics is given by \cite{Oshikawa}
\begin{eqnarray}
  H_{\rm eff} &=& \int \frac{{\rm d}x}{2\pi}\left[
  v_{S}K(\pi\Pi)^{2} + \frac{v_{S}}{K}\left(
    \frac{{\rm d}\phi}{{\rm d}x}\right)^{2}\right] 
  \nonumber \\
  && + v_{S}y_{\phi}\int \frac{{\rm d}x}{2\pi}\cos\sqrt{2}\phi,
\label{eq:sg}
\end{eqnarray}
where $\Pi$ is the momentum density conjugate to $\phi$, 
$[\phi(x),\Pi(x')] = i\delta(x-x')$, 
$v_{S}$ is the sound velocity of the system. 
The dual field $\theta$ is defined as $\partial_{x}\theta = \pi\Pi$. 
The correlation exponents of several operators are governed by the coupling 
$K$ and the scaling dimension of the operator $\sqrt{2}\cos\sqrt{2}\phi$ 
is $x=K/2$. 
Thus the second term in the Hamiltonian (\ref{eq:Hamiltonian}) 
is relevant for $K<4$ and irrelevant for $K>4$. 

The one-loop renormalization group equations of this model are given by 
\[
  \frac{{\rm d}K(L)^{-1}}{{\rm d}\ln L}=\frac{1}{8}y_{\phi}^{2}(L),
  \hspace{5mm}
  \frac{{\rm d}y_{\phi}(L)}{{\rm d}\ln L} 
  = \left(2-\frac{K(L)}{2}\right)y_{\phi}(L),
\]
where $L$ is an infrared cutoff. 
These are the recursion relations of Kosterlitz. \cite{Kosterlitz74} 
The renormalization group flow is shown in Figure 1. 
In the region A, the coupling $K$ renormalized to a finite value 
and $y_{\phi}$ to zero;
the system is critical and this corresponds to the no-plateau region 
in the spin system. 
In the regions B and B', $K$ renormalized to $0$ and $|y_{\phi}|$ to 
$\infty$; these regions are massive phases and the field $\phi$ 
is locked to $\pi$ and $0$ for B and B' respectively. 
There is a separatrix 
$32K^{-1}-8\ln K^{-1} -y_{\phi}^{2} = 8 + 8\ln 4$ which separates the infrared 
stable region (A) from the infrared unstable regions (B and B'). 
On this separatrix the BKT transition takes place. 
Near $(K,y_{\phi})=(4,0)$, and with the notation $K=4(1+y_{0}/2)$, 
the BKT transition point is on the line $|y_{\phi}| = y_{0}$ and we obtain
\begin{equation}
  y_{0}(L) = \frac{y_{0}}{y_{0}\ln (L /L_{0})+1}, 
\end{equation}
where $y_{0}$ and $L_{0}$ are the bare values. 


According to the numerical calculation 
by Oshikawa {\it et al.} \cite{Oshikawa} for the $S=3/2$ Heisenberg chain 
($D=0$) at the magnetization $M=M_{S}/3$, 
this Gaussian coupling $K$ is $K\approx4.4$, 
(in the notation of ref. \cite{Oshikawa} 
the compactification radius is given by $R=(2\pi K)^{-1/2}=0.95/\sqrt{8\pi}$), 
and the system is gapless (no plateau). 
This value is slightly larger than 
the BKT critical (fixed point) coupling $K=4$.  

\section{Numerical approach and results}
\subsection{Finite size behavior}
In order to study numerically, let us consider the finite-size 
behavior of the effective model (\ref{eq:sg}) based on the 
conformal invariance.\cite{Cardy} 
The scaling dimension of the primary field 
${\cal{O}}_{m,n}=\exp ({\rm i} m\sqrt{2}\phi+ {\rm i} n\sqrt{2}\theta)$ 
for the fixed point Hamiltonian ($y_{\phi}=0$) 
with the periodic boundary condition (PBC) is given by 
\begin{equation}
  x_{m,n} = \frac{K}{2}m^{2} + \frac{1}{2K}n^{2}, 
\end{equation}
where $m$ and $n$ are the ``magnetic'' and the ``electric'' charges 
in the Coulomb gas.\cite{Kadanoff}  
According to the conformal field theory, 
the state of the finite-size system have a correspondence 
to the operator in the infinite-size system, and 
the excitation energy of the finite-size system is given by 
\begin{equation}
  \Delta E_{m,n}(L)=E_{m,n}(L) -E_{g}(L) = \frac{2\pi v_{S}}{L}x_{m,n}, 
\label{eq:coulomb}
\end{equation}
where $E_{g}(L)$ is the 
ground state energy of $L$-site with PBC. 

The BKT transition takes place between the gapless and gapful regions, 
and the PRG analysis is affected 
by several irrelevant corrections sensitively. 
Moreover, on the transition point, there exists a logarithmic size correction 
in the excitation energy. 
PRG analysis does not suppose these features. 
In order to determine the BKT transition point, we apply the method 
proposed by Nomura and Kitazawa \cite{Nomura98} 
(see also refs \cite{Okamoto} and \cite{Kitazawa99}), 
which is free from the logarithmic size correction. 
If we can have the half-integer for the ``magnetic'' charge $m$, 
which is not permitted for PBC by 
the translational symmetry, 
we have the following finite size spectrum for $m=\pm 1/2$, 
\begin{eqnarray}
   x_{\pm 1/2,0}^{c}(L)  &\equiv& 
   \frac{L}{2\pi v_{S}}\Delta E_{\pm 1/2,0}^{c}(L) =
   \frac{K(L)}{8}+\frac{1}{2}y_{\phi}(L), \nonumber \\
   x_{\pm 1/2,0}^{s}(L) &\equiv& 
   \frac{L}{2\pi v_{S}}\Delta E_{\pm 1/2,0}^{s}(L) =
   \frac{K(L)}{8}-\frac{1}{2}y_{\phi}(L),
\label{eq:half}
\end{eqnarray}
where the energies $\Delta E_{\pm 1/2,0}^{c,s}(L)$ correspond to 
the operators $\sqrt{2}\cos\phi/\sqrt{2}$ and $\sqrt{2}\sin\phi/\sqrt{2}$ 
respectively. 
The term with $y_{\phi}$ comes from the first order 
perturbation and from the operator product expansions,\cite{Kitazawa}
\begin{eqnarray}
  \sqrt{2}\cos \sqrt{2}\phi\left[\sqrt{2}\cos \frac{1}{\sqrt{2}}\phi\right]
  &\sim& \sqrt{2}\cos \frac{1}{\sqrt{2}}\phi,
  \nonumber \\
  \sqrt{2}\cos \sqrt{2}\phi\left[\sqrt{2}\sin \frac{1}{\sqrt{2}}\phi\right]
  &\sim& \sqrt{2}\sin \frac{1}{\sqrt{2}}\phi. \nonumber
\end{eqnarray}
Near the BKT transition point $|y_{\phi}|=y_{0}$ ($K=4(1+y_{0}/2)$), 
defining $y_{\phi} = \pm y_{0}(1+t)$ where $t$ measures the distance 
from the transition point, we have for $y_{\phi}>0$ 
\begin{eqnarray}
  x_{\pm 1/2,0}^{c}(L) &=&
  \frac{1}{2} + \frac{3}{4}y_{0}(L)\left(1+\frac{2}{3}t\right), \nonumber \\
  x_{\pm 1/2,0}^{s}(L) &=& \frac{1}{2} 
     - \frac{1}{4}y_{0}(L)\left(1+2t\right),
\label{eq:phidp}
\end{eqnarray}
and for $y_{\phi}<0$, 
\begin{eqnarray}
  x_{\pm 1/2,0}^{c}(L) &=& \frac{1}{2} 
     - \frac{1}{4}y_{0}(L)\left(1+2t\right), \nonumber \\
  x_{\pm 1/2,0}^{s}(L) &=& \frac{1}{2} 
     + \frac{3}{4}y_{0}(L)\left(1+\frac{2}{3}t\right).
\label{eq:phidm}
\end{eqnarray}
On the other hand, for $(m,n)=(0,\pm2)$ we have 
\begin{equation}
  x_{0,\pm 2}(L) \equiv \frac{L}{2\pi v_{S}}\Delta E_{0,\pm 2}(L)
   = \frac{2}{K} = \frac{1}{2} -\frac{1}{4}y_{0}(L).
\label{eq:thetad}
\end{equation}
From eqs. (\ref{eq:phidp}), (\ref{eq:phidm}) and (\ref{eq:thetad}), 
energy differences $\Delta E_{\pm 1/2,0}^{s(c)}(L)$ and $\Delta E_{0,\pm 1}$ 
cross linearly at the BKT transition point $t=0$ for $y_{\phi} >0$ 
$(y_{\phi}<0)$, and this behavior can be used 
to determine the BKT critical point.

From (\ref{eq:half}), $\Delta E_{\pm 1/2,0}^{c}$ 
and $\Delta E_{\pm 1/2,0}^{s}$ cross at $y_{\phi}=0$.\cite{Kitazawa} 
This is the Gaussian fixed point. For $K<4$, 
the operator $\sqrt{2}\cos\sqrt{2}\phi$, 
which is in the second term of the Hamiltonian (\ref{eq:sg}), 
is relevant and a second order phase transition occurs at this point. 
On this transition point, we have 
$\Delta E_{\pm 1/2,0}^{c,s}<\Delta E_{0,\pm 2}$. 

The ``electric charge'' $n$ in eq. (\ref{eq:coulomb}) relates to the 
variation of the magnetization from $M=M_{S}/3$, and 
the excitation energy $\Delta E_{0,\pm 2}$ 
in the spin system is described as
\begin{eqnarray}
  \lefteqn{\Delta E_{0,\pm 2}(L)} \nonumber \\
  &=& \frac{E(M_{S}/3+2,L)+E(M_{S}/3-2,L)-2 E(M_{S}/3,L)}{2},
\end{eqnarray}
where $E(M,L)$ is the lowest energy for the magnetization $M$ 
with the PBC. 
The excitation energies corresponding to the operator 
$\sqrt{2}\cos\phi/\sqrt{2}$ and $\sqrt{2}\sin\phi/\sqrt{2}$ 
are obtained by the two lowest energies with the twisted boundary 
condition (TBC) \cite{Blote,Alcaraz,Destri,Fukui} 
$S_{L+1}^{\pm}=-S_{1}^{\pm}$, $S_{L+1}^{z}=S_{1}^{z}$ as 
\begin{eqnarray}
  \Delta E_{\pm 1/2,0}^{c}(L)
  &=& E^{\rm TBC}(M_{S}/3,L,1) - E(M_{S}/3,L) \\
  \Delta E_{\pm 1/2,0}^{s}(L) &=& E^{\rm TBC}(M_{S}/3,L,-1) - E(M_{S}/3,L)
\nonumber
\end{eqnarray}
where $E^{TBC}(M_{S}/3,L,P)$ is the lowest energy with the magnetization 
$M=M_{S}/3$ and the parity 
$P: \mbox{\boldmath$S$}_{j}\rightarrow \mbox{\boldmath$S$}_{L-j+1}$. 

\subsection{Numerical results}
Here we show our numerical results. 
We calculate the energy of finite-size systems $L=8$, $10$, $12$, $14$ 
with the numerical diagonalization calculation. 
In the region $D>0$, the lowest energy state with the magnetization 
$M$ and with PBC has the parity $P=(-1)^{M+L/2}$ and the momentum 
$q=(M+L/2)\times \pi$. 

Figure 2 shows the energy differences $\Delta E_{\pm1/2,0}^{c}$, 
$\Delta E_{\pm1/2,0}^{s}$, and $\Delta E_{0,\pm 2}$ for $L=14$ systems. 
We find that among these three energies 
the lowest one is $\Delta E_{0,\pm 2}$ for small $D$ ($0<D<D_{c_{1}}$) 
and $\Delta E_{\pm 1/2,0}^{c}$ for large $D$ ($D_{c2} < D$). 
In the intermediate region $D_{c1}< D<D_{c2}$, 
 $\Delta E_{\pm 1/2,0}^{s}$ is lower than $\Delta E_{\pm 1/2,0}^{c}$ and 
$\Delta E_{0,\pm 2}$. 
In Figure 1, the small $D$ region $0<D<D_{c1}$ 
should correspond to the region A, 
the intermediate region $D_{c1}<D<D_{c2}$ to B, 
and the large $D$ region to B'. 

From the critical properties of the effective model 
(\ref{eq:sg}) [see eqs. (\ref{eq:phidp}) and (\ref{eq:thetad})], 
the point $D_{c1}$ should be a BKT critical point. 
The size dependence of the crossing points between $\Delta E_{\pm 1/2,0}^{s}$ 
and $\Delta E_{0,\pm 2}$ is shown in Figure 3 and the extrapolated 
value is given by $D_{c1}=0.387$. 
For small $D$ ($<D_{c1}$) the excitation spectrum is gapless and 
there does not exist a plateau in the magnetization curve. 

In order to check this, we calculate 
the averaged scaling dimension,\cite{Nomura98,Okamoto,Kitazawa99} 
\begin{equation}
  \left( x_{\pm 1/2, 0}^{c}(L) +3x_{\pm 1/2, 0}^{s}(L)\right)/4,
\label{eq:dmnsnb}
\end{equation}
at the point $D=D_{c1}$. 
From (\ref{eq:phidp}), 
this value cancels the leading logarithmic size correction, 
and should be $1/2$ at $D=D_{c1}$. 
To calculate the scaling dimension, we need the sound velocity $v_{S}$. 
This can be calculated using the lowest energy with $M=M_{S}/3$ and the 
wave-number $q=2\pi /L$ (corresponding to the U(1) current), 
\begin{equation}
  v_{S}(L) = \frac{E(M_{S}/3, L,q=2\pi /L)-E(M_{S}/3,L)}{2\pi /L}. 
\end{equation}
We extrapolate this finite size value in the polynomial of $1/L^{2}$, and 
obtain $2\pi v_{S}=20.06$. 
Figure 4 shows the size dependence of the value (\ref{eq:dmnsnb}). 
We can see that the extrapolated value is very close to 
the expected one $1/2$. 
We also calculate the conformal anomaly number $c$ from 
\begin{equation}
  \frac{E(M_{S}/3,L)}{L} = \epsilon_{m=1/2} - \frac{\pi v_{S}c}{6L^{2}}
  +\cdots,
\label{eq:anomalyN}
\end{equation}
where $\epsilon_{m=1/2}$ is the energy per site of the limit 
$L\rightarrow \infty$, $c$ is the conformal anomaly number, and 
$\cdots$ means the higher size correction. 
We estimate $c$ at $D=D_{c1}$ as $c=0.97$, which is close to the 
expected value $c=1$. 

Thus we can conclude that the point $D=D_{c1}$ is the BKT critical point. 
The width of the plateau near the transition point behaves singularly as 
$\Delta h \sim \exp (-\mbox{const.}/\sqrt{D-D_{c1}})$. 

According to (\ref{eq:half}), 
the crossing point $D_{c2}$ between $\Delta E_{\pm 1/2,0}^{c}$ and 
$\Delta E_{\pm 1/2,0}^{s}$ should be a second order phase transition point 
($K<4$ and $y_{\phi}=0$). 
For large $D$ ($D_{c2}< D$), the system is in the large-$D$ region and a 
plateau appears in the magnetization curve at $M=M_{S}/3$. 
Thus we have another massive phase in the region $D_{c1}<D<D_{c2}$. 
As a possibility, this intermediate region should be characterized 
by the following partially magnetized valence-bond-solid (VBS) state, 
\begin{eqnarray}
\lefteqn{|\mbox{mVBS},\mbox{PBC}\rangle}\nonumber \\
 &=&a_{L}^{\dagger}(a_{L}^{\dagger}b_{1}^{\dagger}-b_{L}^{\dagger}a_{1}^{\dagger})
|\prod_{j=1}^{L-1}a_{j}^{\dagger}
(a_{j}^{\dagger}b_{j+1}^{\dagger}-b_{j}^{\dagger}a_{j+1}^{\dagger})|0\rangle. 
\end{eqnarray}
where we describe the spin state by the Schwinger bosons, 
that is, $a_{j}^{\dagger}$ ($b_{j}^{\dagger}$) creates the 
$S=1/2$ $\uparrow$ ($\downarrow$) spin at the $j$-th site, and 
$|0\rangle$ is the vacuum of the Schwinger bosons. Here we assume PBC. 
This partially magnetized VBS state had firstly been discussed 
by Oshikawa\cite{Oshikawa92} 
for another model.(see also ref. \cite{Oshikawa}) 

Let us consider the crossing behavior near the point $D=D_{c2}$ 
on another point of view. 
The large-$D$ phase is characterized by the state
\begin{equation}
 |\mbox{large-}D\rangle 
  = \prod_{j=1}^{L}a_{j}^{\dagger}(a_{j}^{\dagger}b_{j}^{\dagger})|0\rangle.
\label{eq:large-D}
\end{equation}
With the twisted boundary condition ($a_{L+1}^{\dagger}=a_{1}^{\dagger}$, 
$b_{L+1}^{\dagger}=-b_{1}^{\dagger}$), 
the magnetized valence-bond-solid state is described as
\begin{eqnarray}
  \lefteqn{|\mbox{mVBS}, \mbox{TBC}\rangle}\nonumber \\
  &=&
a_{L}^{\dagger}(a_{L}^{\dagger}b_{1}^{\dagger}+b_{L}^{\dagger}a_{1}^{\dagger})
\prod_{j=1}^{L-1}a_{j}^{\dagger}
(a_{j}^{\dagger}b_{j+1}^{\dagger}-b_{j}^{\dagger}a_{j+1}^{\dagger})|0\rangle, 
\end{eqnarray}
while the large-$D$ state is also characterized by (\ref{eq:large-D}). 
The state $|\mbox{mVBS}, \mbox{TBC}\rangle$ has 
the parity of the space inversion $P=-1$, 
while the large-$D$ state has $P=1$. 
This explains the level crossing between $\Delta E_{\pm 1/2,0}^{c}$ 
and $\Delta E_{\pm 1/2,0}^{s}$ in Figure 2. 

The size dependence of the crossing point is shown in Figure 5. 
The estimated critical point between the intermediate 
and the large-$D$ regions is $D_{c}=0.943$, which is very close to 
the value $D_{c}=0.93\pm 0.01$ obtained by Sakai and Takahashi. 
However, the phase transition is not of the BKT type 
but a second order phase transition 
which is the same universality class as the Haldane - large-$D$ transition in 
the $S=1$ anisotropic model,\cite{Botet,Solyom,Schulz} 
\begin{equation}
  H_{S=1} = \sum_{j=1}^{L}[S_{j}^{x}S_{j+1}^{x}+S_{j}^{y}S_{j+1}^{y}
  +\Delta S_{j}^{z}S_{j+1}^{z}]
  +D\sum_{j=1}^{L} (S_{j}^{z})^{2}.
\label{eq:S=1}
\end{equation}
Especially, the intermediate region of the model (\ref{eq:Hamiltonian}) 
corresponds to the $S=1$ Haldane phase in (\ref{eq:S=1}). 

To verify this speculation, we calculate the coupling $K$. 
The sound velocity is evaluated as $2\pi v_{S}=21.87$. 
Using this value, we determine $K$ from $x_{\pm 1/2,0}^{s}(L)$, 
$x_{\pm 1/2,0}^{c}(L)$ [eqs. (\ref{eq:half})], and 
\begin{equation}
  \frac{L}{2\pi v_{S}}\Delta E_{0,\pm 1}(L) 
  = x_{0,\pm 1}(L) = 1/2K,
\end{equation}
\begin{eqnarray}
  \lefteqn{\Delta E_{0,\pm 1}(L)} \nonumber \\
  &=& \frac{E(M_{S}/3+1,L)+E(M_{S}/3-1,L)-2 E(M_{S}/3,L)}{2}, \nonumber
\end{eqnarray}
as shown in Figure 6. 
Extrapolated values of these three are consistent and we obtain $K=3.77$. 
This is smaller than the BKT fixed point value, and the point $D_{c2}=0.943$ 
should be a second order phase transition point. 
We also estimate the conformal anomaly number form eq. (\ref{eq:anomalyN}), 
and check that it is very close to the expected value $c=1$. 

In the vicinity of the critical point $D=D_{c2}$, 
the width of the magnetization plateau decreases as 
$\Delta h \sim |D-D_{c2}|^{\nu}$, where the critical exponent $\nu$ is 
given by $\nu = 1/\left( 2 -x_{\pm 1,0}\right) = 8.68$ 

\section{Summary and discussion}
We studied the $S=3/2$ spin chain with the single ion anisotropy 
in a magnetic field (\ref{eq:Hamiltonian}). 
With the numerical approach based on the field theoretical analysis, 
we found a new region between the no-plateau (small $D$) and 
the magnetized large-$D$ (large $D$) regions. 
As far as we know, this is the first work in which no-plateau state and
two kinds of plateau states are found when the parameter is changed
in a simple and realistic model.
This intermediate region should have the same character 
of the partially magnetized valence-bond-solid state, 
and should be characterized 
by the string order parameter\cite{Oshikawa92,Nijs,Tasaki} 
and by the edge states for the open boundary system\cite{AKLT,Kennedy90} 
reflecting a hidden discrete symmetry.\cite{Oshikawa92,Kennedy92} 
But in our small size calculation, we could not collect evidence, 
which remains for future study. 

Restricting to three states $S^{z}=3/2$, $1/2$, and $-1/2$, 
(in sufficiently large magnetic field)
Sakai and Takahashi mapped the $S=3/2$ model 
to an $S=1$ generalized anisotropic model with $h=0$. 
In order to investigate the relation between the $S=1$ with $h=0$ 
and the $S=3/2$ model in a magnetic field, 
we further applied the same analysis to the following model, 
\begin{eqnarray}
  H_{\Delta}&=&\sum_{j}^{L}(S_{j}^{x}S_{j+1}^{x}+S_{j}^{y}S_{j+1}^{y}
  +\Delta S_{j}^{z}S_{j+1}^{z}) \nonumber \\
  && +D\sum_{j}^{L}(S_{j}^{z})^{2} -h \sum_{j}^{L}S_{j}^{z},
\end{eqnarray}
in the region $0\leq \Delta\leq 1$. 
Figure 7 shows the energy differences $\Delta E_{\pm 1/2,0}^{c}$, 
$\Delta E_{\pm 1/2,0}^{s}$, $\Delta E_{0,\pm2}$ for $L=14$, $\Delta=0.5$ 
system. 
The lowest energy difference is $\Delta E_{0,\pm 2}$ for small $D$ 
and $\Delta E_{\pm 1/2,0}^{c}$ for large $D$. 
In this case, there is a transition between the no-plateau and the 
large-$D$ regions. 
[The crossing point of $\Delta E_{\pm 1/2,0}^{c}$ and 
$\Delta E_{\pm 1/2,0}^{s}$ is the Gaussian fixed point $y_{\phi}=0$ 
in the no-plateau region ($K>4$).]

We show the $\Delta$-$D$ diagram in Figure 8. 
We found that for $0.808 < \Delta<1$, the $M=M_{S}/3$ phase structure 
is the same as is of the $\Delta =1$ case 
and there exists the intermediate magnetic region ($D_{c1}<D<D_{c2}$). 
But for $0<\Delta <0.808$, the direct transition (BKT type) 
occurs between the no-plateau and the large-$D$ regions. 
This boundary runs from $(\Delta,D) = (0,1.312)$ to 
the multicritical point $(0.808,0.817)$. 
The phase structure is topologically the same as is 
of the $S=1$ anisotropic model \cite{Botet,Solyom,Schulz}
with $h=0$ (\ref{eq:S=1}). 

We acknowledge K. Nomura, M. Oshikawa, and T. Sakai for discussions. 
The computation in this work was performed using the facilities of the 
Supercomputer Center, Institute for Solid State Physics, University of 
Tokyo.

\begin{figure}[h]
\begin{center}
\leavevmode \epsfxsize=3.2in \epsfbox{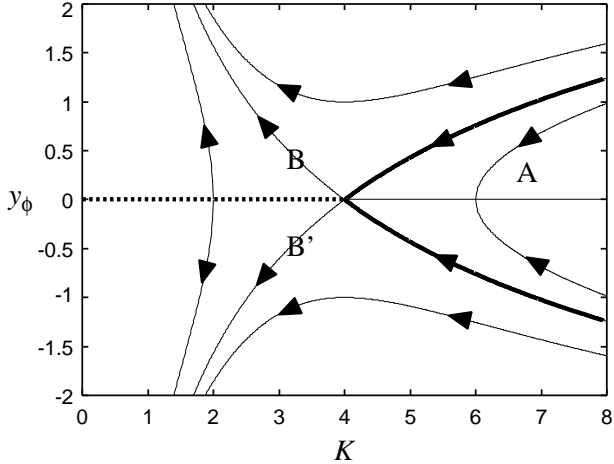}
\end{center}
\caption{Renormalization group flow diagram. 
The bold lines (separating the regions A and B, and A and B') 
are the BKT critical lines. 
On the dotted line between the regions B and B' ($y_{\phi}=0$, $K<4$), 
a second order phase transition occurs.}
\end{figure}

\begin{figure}[h]
\begin{center}
\leavevmode \epsfxsize=3.2in \epsfbox{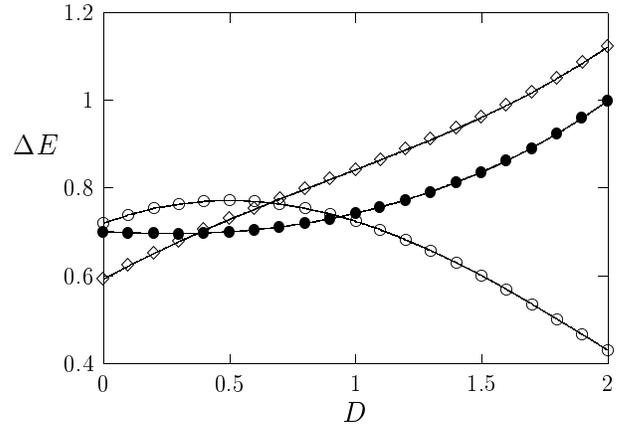}
\end{center}
\caption{Energy differences $\Delta E_{\pm 1/2,0}^{c}$ ($\circ$), 
$\Delta E_{\pm 1/2,0}^{s}$ ($\bullet$), 
and $\Delta E_{0,\pm 1}$ ($\Diamond$) for the $L=14$ system. 
The crossing point of $\Delta E_{\pm 1/2,0}^{s}$ and $\Delta E_{0,\pm 2}$ 
is the BKT critical point. 
The crossing point of $\Delta E_{\pm 1/2,0}^{c}$ 
and $\Delta E_{\pm 1/2,0}^{s}$ 
is a second order phase transition point. }
\end{figure}

\begin{figure}[h]
\begin{center}
\leavevmode \epsfxsize=3.2in \epsfbox{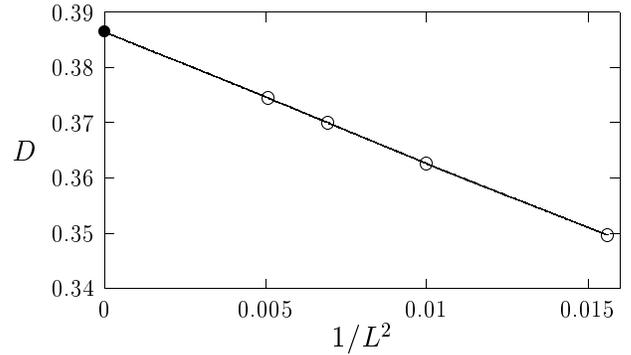}
\end{center}
\caption{Size dependence of the crossing point between the energy differences 
$\Delta E_{\pm 1/2,0}^{s}$ and $\Delta E_{0,\pm 2}$. 
The BKT critical point is estimated as $D_{c1}=0.387$. }
\end{figure}

\begin{figure}[h]
\begin{center}
\leavevmode \epsfxsize=3.2in \epsfbox{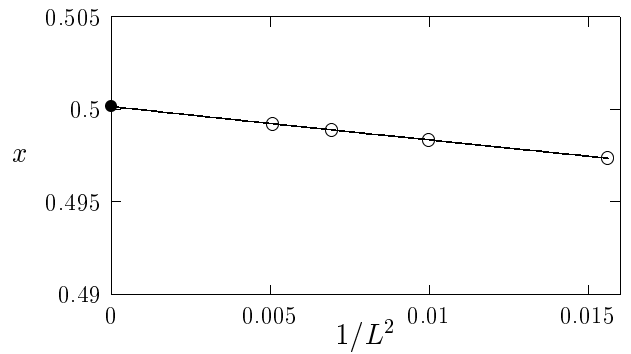}
\end{center}
\caption{Size dependence of the averaged scaling dimension 
$(x_{\pm 1/2,0}^{c}+3x_{\pm 1/2,0}^{s})/4$ at the BKT critical point 
$D_{c1}=0.387$.}
\end{figure}

\begin{figure}[h]
\begin{center}
\leavevmode \epsfxsize=3.2in \epsfbox{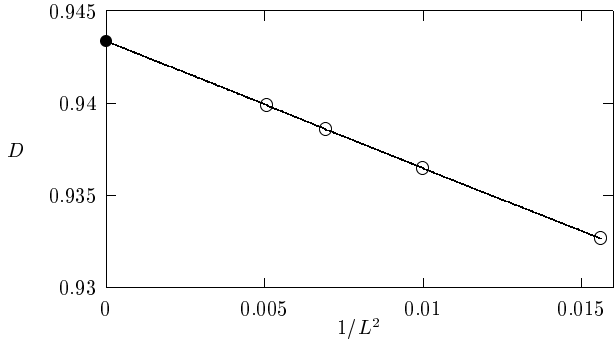}
\end{center}
\caption{Size dependence of the crossing point between the energy differences 
$\Delta E_{\pm 1/2,0}^{c}$ and $\Delta E_{\pm 1/2,0}^{s}$. 
The critical point between the intermediate 
and the magnetic large-$D$ regions is estimated as $D_{c2}=0.943$.}
\end{figure}

\begin{figure}[h]
\begin{center}
\leavevmode \epsfxsize=3.2in \epsfbox{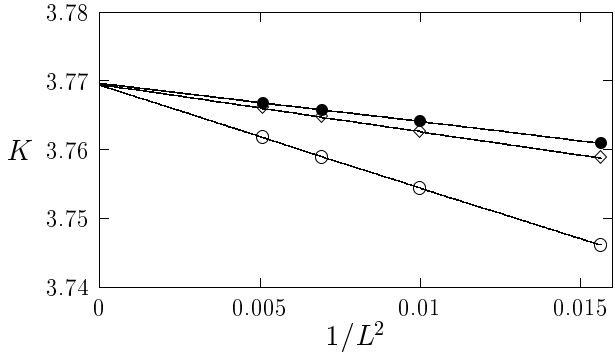}
\end{center}
\caption{The parameter $K$ obtained by 
the data $x_{\pm 1/2,0}^{c}=K/8$ ($\circ$), 
$x_{\pm 1/2,0}^{s}=K/8$ ($\bullet$), and $x_{0,\pm 1}=1/2K$ ($\Diamond$). 
The extrapolated value is $K=3.77$.}
\end{figure}

\begin{figure}[h]
\begin{center}
\leavevmode \epsfxsize=3.2in \epsfbox{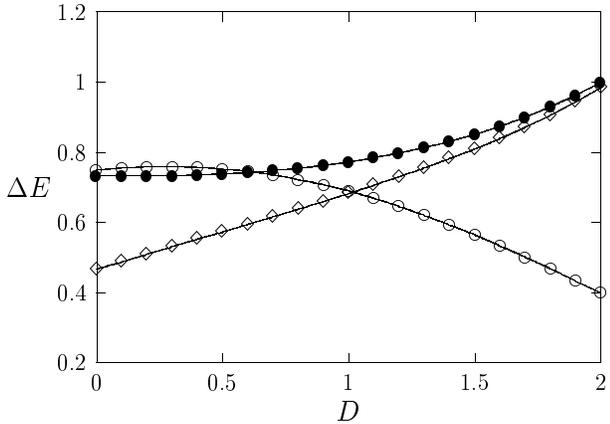}
\end{center}
\caption{Energy differences $\Delta E_{\pm 1/2,0}^{c}$ ($\circ$), 
$\Delta E_{\pm 1/2,0}^{s}$ ($\bullet$), 
and $\Delta E_{0,\pm 2}$ ($\Diamond$) for the system 
with $\Delta=0.5$ and $L=14$. 
The crossing point of $\Delta E_{\pm 1/2,0}^{c}$ and 
$\Delta E_{0,\pm 2}$ is the boundary between the no-plateau and the large-$D$ 
regions. 
The crossing point of $\Delta E_{\pm 1/2,0}^{c}$ and 
$\Delta E_{\pm 1/2,0}^{s}$ is the Gaussian fixed point in the no-plateau 
region.}
\end{figure}
\begin{figure}[h]
\begin{center}
\leavevmode \epsfxsize=3.2in \epsfbox{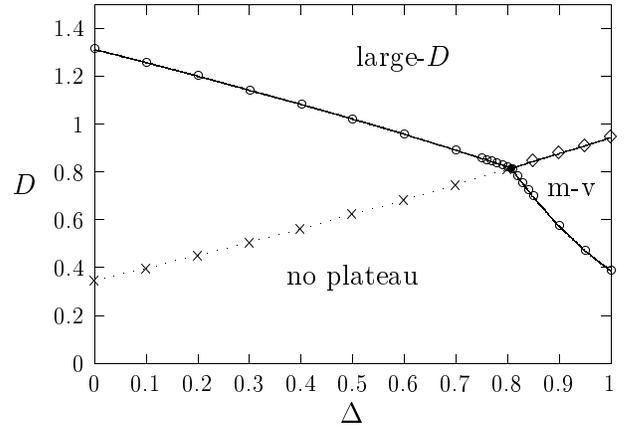}
\end{center}
\caption{The boundary of the no-plateau, the large-$D$, 
and the magnetized VBS (m-v) regions. 
The boundary between the no-plateau and the large-$D$ regions, 
and between the no-plateau and the magnetized VBS regions ($\circ$) 
are the BKT critical lines. 
The boundary between the large-$D$ and the magnetized VBS regions ($\Diamond$)
is a second order critical line. 
The multicritical point of these regions ($\bullet$) are estimated as 
$(\Delta,D)=(0.808,0.817)$. 
We also show the Gaussian fixed line in the no-plateau region ($\times$).}
\end{figure}

\end{document}